# Geothermal Energy in Sedimentary Basins: Assessing Techno-economic Viability for Sustainable Development

**Orkhan Khankishiyev[1], Saeed Salehi[1], Cesar Vivas[1], Runar Nygaard[1], and Danny Rehg[2]**

**[1] Well Construction Technology Center - The University of Oklahoma, Norman, OK, USA.**

**[2] Criterion Energy Partners, Houston, TX, USA**

## ABSTRACT

Enhanced Geothermal Systems (EGS) models operating in super-hot igneous rocks have demonstrated significantly improved heat transfer rates and power production compared to conventional geothermal systems. However, the drilling of deep geothermal wells has proven to be a challenging endeavor, primarily due to issues such as loss circulation events, material limitations under high temperatures, and the production of corrosive fluids. Furthermore, the substantial upfront costs, coupled with geological and technical obstacles associated with drilling super-hot EGS wells in igneous rocks, hinder the widespread implementation of geothermal systems.

Alternatively, geothermal energy development in sedimentary basins presents an opportunity for clean energy production with relatively lower investment costs compared to the development of super-hot EGS in igneous rocks. Sedimentary basins exhibit attractive temperatures for geothermal applications, and their wide distribution enhances the potential for nationwide deployment. Decades of drilling and development experience in oil and gas wells have yielded a wealth of data, knowledge, and expertise. Leveraging this experience and data for geothermal drilling can significantly reduce costs associated with subsurface data gathering, well drilling, and completion.

This paper explores the economic viability of geothermal energy production systems in sedimentary basins. The study encompasses an analysis of time-to-hit-temperature (THT) and cost-to-hit-temperature (CHT) parameters, as well as Favorability maps across the United States. These maps are based on factors such as well depth, total drilling time, well cost, and subsurface temperature data. By integrating sedimentary basin maps and underground temperature maps, the THT and CHT maps can facilitate the strategic placement of EGS wells and other geothermal system applications in the most favorable locations across the United States.

## 1. Introduction

Despite being one of the low-cost sources for energy production, the utilization of geothermal energy accounts for only 0.24% of the total energy consumption. In contrast, petroleum and natural gas comprise a staggering 68% of the total energy consumption in the United States, according to the U.S. Energy Information Agency(EIA, 2022). Geothermal energy systems can be primarily categorized into three groups: direct use and district heating systems, geothermal heat pumps, and geothermal power plants, Tester et al. (2006). The limited availability of naturally occurring hydrothermal reservoirs is often cited as a key factor contributing to the relatively small contribution of geothermal energy production, Gurgenci et al. (2008). However, investigations into the geothermal potential of the Earth have revealed that substantial amounts of energy are stored at reachable depths, ranging up to 10 km. Considering the distribution of sedimentary basins in the United States (EIA, 2015), a significant amount of thermal energy is stored in sedimentary rocks, which have been subject to exploration for several decades. Tester et al. (2006) estimated the recoverable thermal energy potential of the United States to be around 5.6*106 EJ, with over 100,000 EJ of that energy being stored in sedimentary rocks. To provide perspective, the annual energy consumption of the United States in 2021 amounted to 92.97 EJ, indicating that the existing recoverable geothermal energy has the potential to meet the nation's energy demands for more than 1000 years.

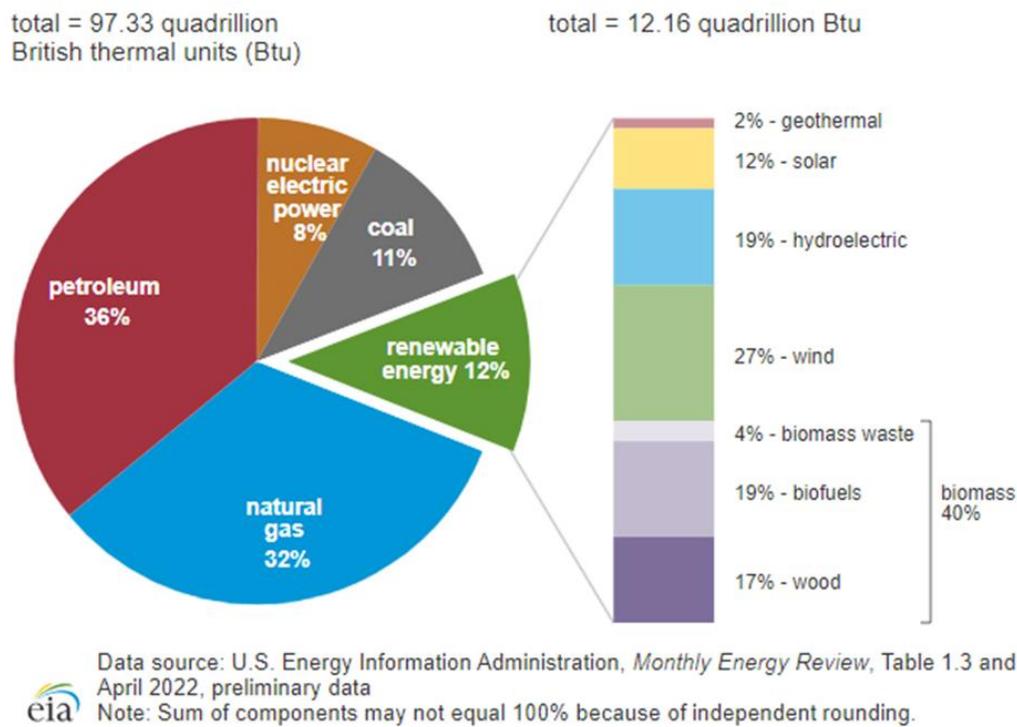

**Figure 1. U.S. primary energy consumption by energy source, 2021,[5]**

The recent upward trend in oil prices, escalating exploration costs of hydrocarbons, the significance of energy security in national economies, and the global push towards achieving net-zero energy production for a sustainable future have drawn attention to geothermal energy. As a result, multiple projects are planned to demonstrate the immense potential of geothermal energy in providing baseload power for the United States. However, achieving such efficiency requires the development of cutting-edge drilling technology and advanced geothermal fluid production systems, along with further exploration of geothermal resources.

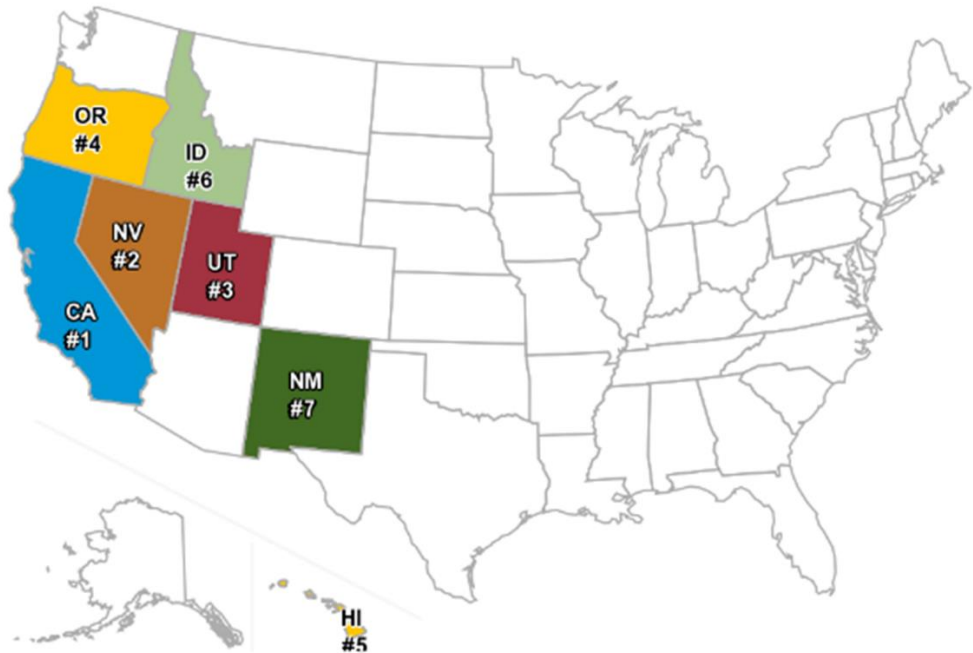

Data source: U.S. Energy Information Administration, *Electric Power Monthly*, Table 16.1.B, February 2022, preliminary data

**Figure 2. State rankings for geothermal electricity generation, 2021, [5]**

In this study, the authors conducted an analysis of the drilling and completion time of 1,074,266 wells drilled in various locations across the United States from 1990 to 2022. The primary focus of this analysis was to investigate the impact of geological parameters on drilling durations. Furthermore, the authors derived the total well cost by employing different correlations that account for factors such as total well depth, well complexity index, and well construction time.

The findings of this analysis facilitated the creation of time to hit temperature (THT) and cost to hit temperature (CHT) maps for the United States. These maps offer valuable insights into the temporal and financial aspects associated with reaching the desired temperature levels during geothermal drilling operations. Additionally, the authors integrated the THT and CHT maps with existing subsurface temperature maps to develop a comprehensive geothermal drilling favorability map for the United States.

By combining the THT and CHT maps with subsurface temperature data, the geothermal drilling favorability map identifies optimal locations for the development of enhanced geothermal systems (EGS) within sedimentary basins. This map serves as a valuable tool for pinpointing the most favorable areas for EGS implementation, aiding in strategic decision-making and facilitating targeted geothermal energy development initiatives.

## 2. Literature Review

The suitability of geothermal systems in sedimentary basins varies depending on multiple factors, including the intended purpose, subsurface temperature availability, proximity to end users, technological considerations, and power plant development costs. Several promising options exist, ranging from shallow installations of heat pumps for area heating and cooling, to closed-loop systems or enhanced geothermal systems (EGS) utilizing idle oil and gas wells for direct heat or electricity generation. Furthermore, deep to super-deep hot reservoirs can be explored for the development of EGS specifically geared towards electricity generation. It is worth noting that the total energy production potential increases as one moves from heat pumps to super-hot EGS applications, albeit with a corresponding rise in installation costs.

For a comprehensive techno-economic investigation, all aspects of geothermal energy system development must be taken into account. The process typically initiates with the drilling of injection and production wells. Upon successful completion, cold working fluid is pumped into the hot geothermal reservoir through injectors, while hot steam is extracted through production wells for surface heat transfer and subsequent power generation (Fox et al., 2013). While the establishment of power plants and grid systems requires substantial investment, the economic viability of geothermal projects heavily relies on drilling expenditures and successful field exploration. As indicated by Blankenship et al. (2005), the overall cost of well drilling and completion constitutes approximately 30-60% of the total expenses incurred in hydrothermal power plant development. It is worth mentioning that drilling costs tend to increase significantly in the case of super-hot EGS well drilling in igneous rocks, primarily due to the harsher drilling conditions encountered in such scenarios.

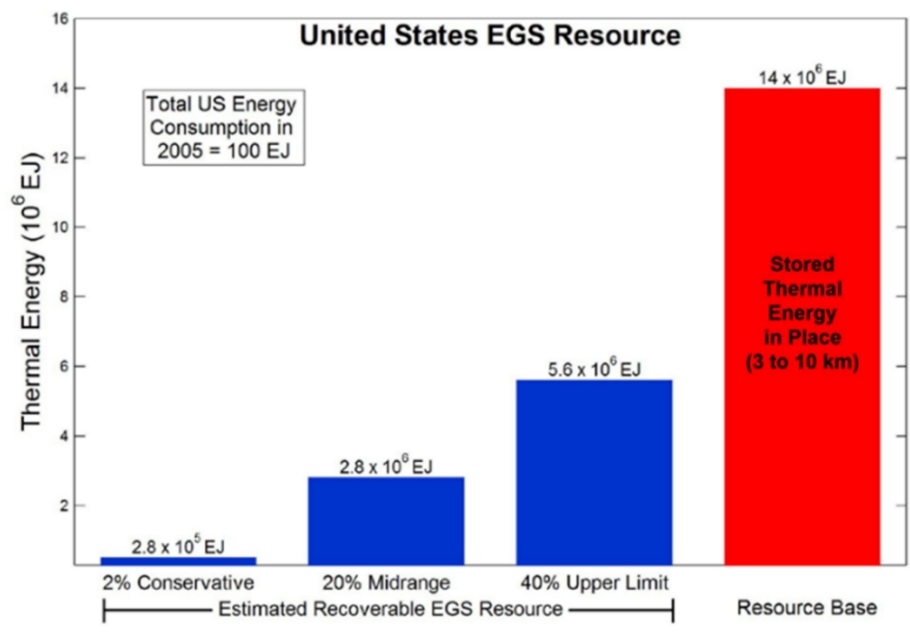

**Figure 3. U.S. Geothermal recourse base up to 10km, (Tester, et al., 2006)**

In a study conducted by Vivas et al. (2020), the technical challenges associated with the development of enhanced geothermal systems (EGS) in super-hot basins were analyzed. On the other hand, Salehi et al. (2022) highlighted several advantages of applying geothermal energy systems in sedimentary basins. Given the higher cost-share and risks involved in geothermal projects, it is crucial to place significant emphasis on the economic analysis of the drilling process, particularly in terms of drilling time and well cost.

While more than 4000 geothermal wells have been drilled, with approximately 3200 of them currently active (Sanyal & Morrow, 2012), the cost of drilling and completion has been rarely disclosed due to the confidential nature of the data. Furthermore, the limited number of geothermal wells drilled across various geological settings makes it challenging to establish a statistically significant trendline for approximating well costs, considering the inherent risks and uncertainties involved. However, it is noteworthy that the drilling and construction processes of hydrocarbon and geothermal wells in sedimentary basins share similarities, enabling the derivation of a relationship between measured depth and well cost based on hydrocarbon well data and its application to geothermal drilling operations.

In a study conducted by Lukawski et al. (2014), the drilling costs of oil and gas wells were evaluated based on data collected from the API Joint Association Survey spanning the years 1976 to 2009. The researchers compared these costs with those associated with geothermal wells. The findings of this comparative analysis revealed that hydrocarbon wells drilled to the same depth exhibited lower costs compared to geothermal wells (Figure 4). To approximate the average drilling and completion costs of geothermal wells, the study proposed an estimation equation with a correlation coefficient of $R^2$=0.92.

$$Geothermal\ Well\ Cost = 1.72 * 10^{-7} * MD^2 + 2.3 * 10^{-3} * MD - 0.62 \quad [6]$$

This equation only uses the measured depth (MD) as an input and does not account for the geologic complexity, well deviation and any other factors affecting the well cost.

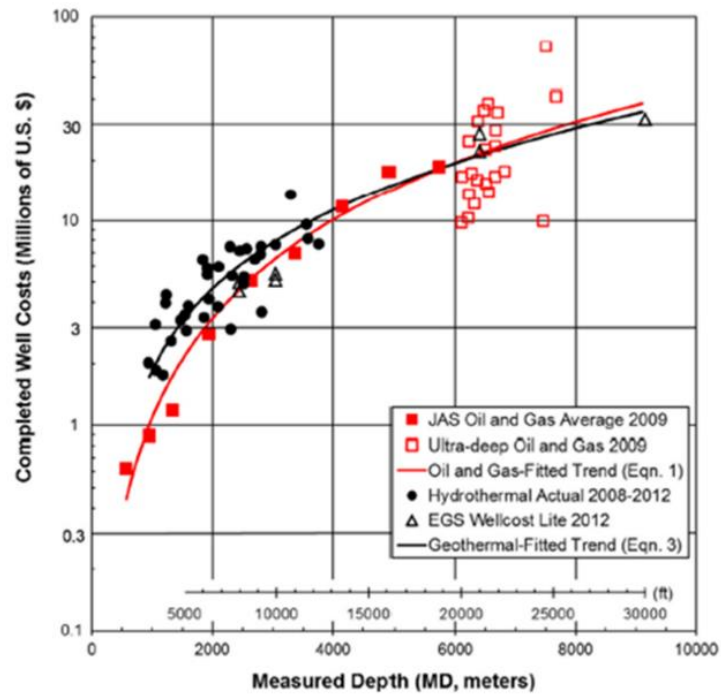

**Figure 4. Geothermal well costs (in black) compared to average 2009 oil and gas well costs (in red). [6]**

In a subsequent study, Lukawski et al. (2016) further expanded upon this uncertainty model by employing probabilistic methods to estimate the distribution of well costs for a range of well depths. By considering the individual cost categories, which delineates the cost breakdown for an EGS well, the researchers developed a more robust framework for evaluating the overall cost uncertainty associated with geothermal wells.

By incorporating these probabilistic approaches and accounting for uncertainties, the aim is to provide more accurate and comprehensive cost estimations for geothermal well drilling and completion. These advancements contribute to a more realistic understanding of the financial aspects and risk factors involved in the development of geothermal energy projects, ultimately supporting better decision-making processes in the industry.

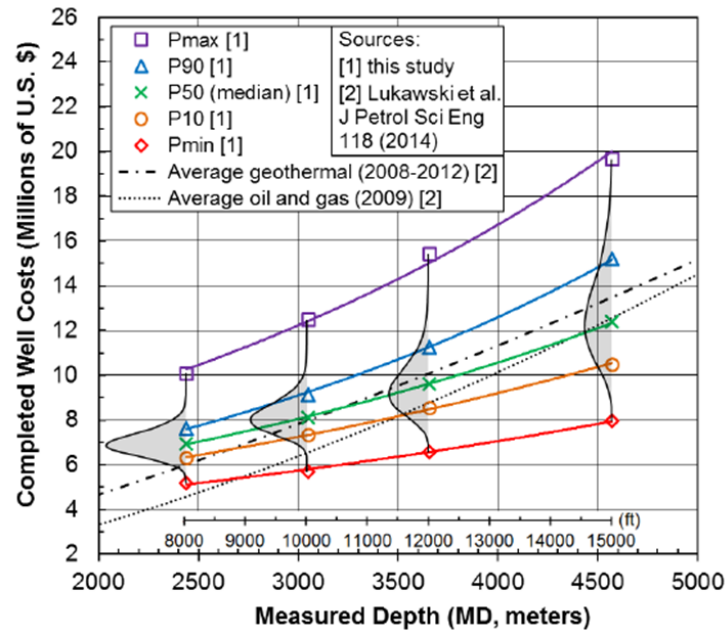

**Figure 5. Probabilistic approach to the well cost trends versus well depth, (Lukawski, et al., 2016)**

In the area of geothermal well cost analysis, Lukawski et al. (2016) conducted a Monte Carlo simulation using data from wells drilled between 2009 and 2013. The simulation revealed that the probability distributions of cost for shallower wells exhibited narrower peaks and less variability, while the opposite was observed for deeper wells. This indicates that deeper wells are relatively rare and more challenging to predict in terms of cost, primarily due to unexpected geological complexities and a higher frequency of operational difficulties.

Contributing to the understanding of underground temperatures in the United States, Blackwell et al. (2011) published an underground temperature map based on a vast dataset from over 35,000 sites. Their analysis considered variables such as thermal conductance, heat flow, and rock density, allowing for the creation of temperature models at various depths. These models serve as valuable resources for assessing non-conventional geothermal resources on a regional to sub-regional scale, providing essential data for geothermal energy exploration and development efforts.

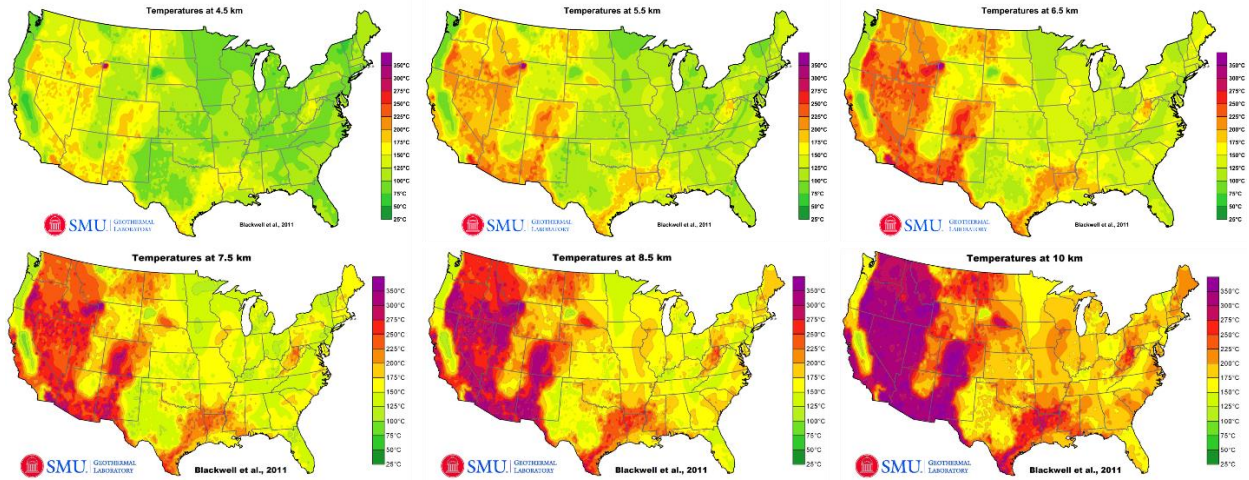

**Figure 6. Temperature-At-Depth Maps for the Conterminous US (Blackwell, et al., 2011)**

Comparing different geological settings, the study revealed that the highest temperatures are predominantly found in basaltic igneous basins. These areas offer promising conditions for geothermal energy extraction. However, it is noteworthy that there are also certain sedimentary basins with elevated temperatures that make them attractive for enhanced geothermal systems (EGS) development. These findings highlight the potential of both igneous and sedimentary basins in supporting geothermal energy initiatives, depending on their respective temperature profiles.

## 2. Methodology

The primary objective of this study is to create Temperature to Hit (THT), Cost to Hit (CHT), and Favorability maps for the United States. These maps aim to visually illustrate the areas where geological and economic factors intersect, creating favorable conditions for the development of Enhanced Geothermal Systems (EGS). The methodology employed in this study is summarized in Figure 7 (Khankishiyev et al. (2023)). The workflow comprises three main parts:

1. Analysis of drilling and completion time: The drilling and completion time of oil and gas wells drilled over the past two decades are analyzed to derive trendlines. This analysis helps establish patterns and trends in drilling and completion times, providing valuable insights for geothermal well development.
2. Estimation of well drilling and completion cost: The well drilling and completion costs are estimated at different fields across the United States. Various factors such as well depth, complexity, and construction time are considered to derive cost estimations. This step provides a comprehensive understanding of the economic aspects associated with geothermal well development.
3. Digitization of temperature maps: The temperature data from Blackwell et al. (2011) temperature maps are digitized. These maps provide information on underground temperatures at different depths and locations across the United States. By digitizing this data, it becomes more accessible and usable for further analysis and integration with other geospatial information.

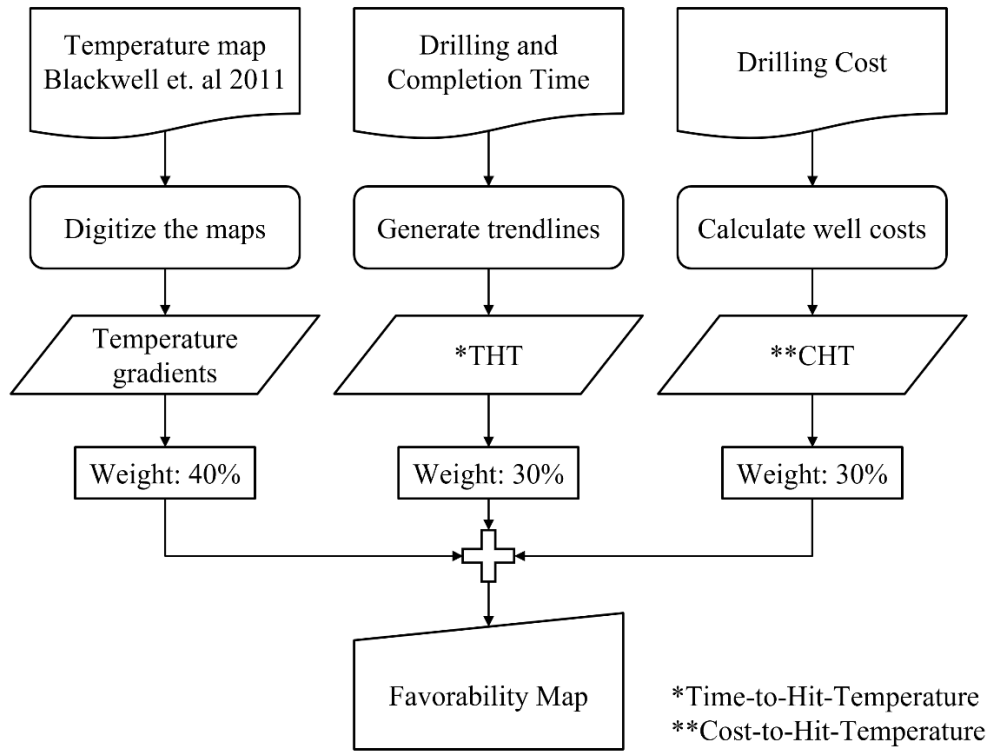

**Figure 7. Flowchart representing the methodology**

By combining the information from these three parts, the study aims to generate THT, CHT, and Favorability maps that highlight the regions where geological conditions and economic feasibility align, indicating favorable locations for the implementation of EGS systems. These maps serve as valuable tools for decision-making and planning in geothermal energy development.

In order to extract temperature data from the Blackwell et al. (2011) temperature maps, a thorough investigation of the methods was conducted. A Matlab code was developed to accurately extract the color codes representing temperatures from the maps. These color codes were then converted into temperature values at specific depths, namely 3.5 km, 4.5 km, 5.5 km, 6.5 km, 7.5 km, 8.5 km, and 10 km.

To further analyze the temperature-depth relationship, temperature-depth gradients were derived for each county individually. This allowed for the determination of temperatures at any given depth within a specific county. Using the derived temperature-depth equation, the depths at which temperatures of 100°C, 150°C, 200°C, 250°C, and 300°C were reached were calculated.

Subsequently, these calculated depth values for different temperature thresholds were inputted into the Geothermal Well Cost equation. This equation considers the relationship between well depth and cost and provides an estimation of the cost required to reach a specific temperature (CHT). By applying this methodology, the study was able to derive the cost associated with achieving desired temperature levels at different depths, providing valuable insights into the economic considerations of geothermal well development.

The drilling and completion time data of wells drilled in 26 U.S. states from 2000 to 2022 underwent a cleaning process to remove outliers using statistical methods. When analyzing the depth-time relationship for these states, it was observed that there was significant scattering, making it challenging to identify a precise trendline to correlate drilling and completion time with total measured depth. To mitigate this issue, linear and exponential trendline equations were derived for the existing 26 states, and the data were averaged at the county level. However, it should be noted that the exponential trendline equations resulted in extremely high values for depths exceeding 23,000 ft (7 km). As a result, the linear trendline equation was selected for extrapolation and prediction of drilling and completion time for pseudo wells based on the depths extracted from the Blackwell et al. (2011) subsurface temperature maps. The resulting values represent the estimated time required to reach the desired temperatures (THT) at different depths.

To generate the favorability maps, a weighted average approach was employed. Well drilling and completion time, cost, and subsurface temperature were assigned weightage factors of 30%, 30%, and 40%, respectively. The data were then normalized on a scale ranging from 1 (least favorable) to 10 (most favorable), allowing for a comprehensive visualization of favorable regions for geothermal energy development. By incorporating these factors and generating favorability maps, the study provides valuable insights into the regions where drilling and completion time, cost, and subsurface temperature align to create favorable conditions for geothermal energy systems. These maps serve as a useful tool for decision-making and strategic planning in geothermal resource exploration and development.

## 3. Results and Discussion

The analysis resulted in the creation of seven distinct maps, including time to hit temperature (THT), cost to hit temperature (CHT), and favorability maps, spanning depths from 3.5 km to 10 km. Each map provides valuable insights into different aspects of geothermal energy development. In the THT maps, varying shades of green represent drilling and completion times of 10 days or less, indicating efficient operations. On the other hand, darker shades of blue indicate longer drilling and completion times, reaching 180 days or more, suggesting more challenging conditions and potential delays. The CHT maps focus on well costs associated with reaching target temperatures. Green shades represent costs of 5 million US dollars or less, indicating relatively affordable drilling and completion processes. As the colors transition to darker shades, costs increase, with the highest range reaching 50 million US dollars or more, indicating higher expenses for achieving desired temperatures.

In contrast to the THT and CHT maps, the favorability map compares different targets at specific depths, as indicated in the map heading. The favorability score is represented by a scale ranging from 1 (light blue) to 10 (blue). Higher scores indicate a more favorable environment for geothermal energy development, considering a combination of factors such as drilling and completion time, cost, and subsurface temperature. These maps provide visual representations of crucial information for decision-making and planning in geothermal energy projects. They allow stakeholders to identify regions with shorter drilling times, lower costs, and higher favorability scores, assisting in the selection of suitable locations for geothermal energy system development. Drilling deeper wells in geothermal energy projects may indeed require more time and incur higher costs. However, the long-term investment return tends to be higher due to the increased geothermal energy potential associated with greater depths.

Certain regions in the United States, such as South to South-East Texas and South Louisiana, exhibit favorable conditions for deep Enhanced Geothermal Systems (EGS) applications. These areas offer relatively faster and more cost-effective drilling operations, along with subsurface temperatures reaching up to 275°C at a depth of 7.5 km. Another example of promising locations for deep EGS applications can be found in North-East Montana and West North Dakota. These regions boast reasonably high geothermal energy potential, with temperatures of around 225°C at approximately 7 km depth. Central to Western Colorado and Eastern Utah are also prime candidates for EGS development. These areas exhibit significantly higher temperatures, surpassing 300°C, starting from a depth of 7 km within sedimentary formations. Such favorable conditions indicate the potential for harnessing substantial geothermal energy resources in these regions. Considering the higher temperatures and the associated energy potential at greater depths, investing in deep EGS projects can prove to be advantageous in the long run. These regions offer a promising outlook for geothermal energy production, considering both the economic and technical aspects of drilling and resource availability.

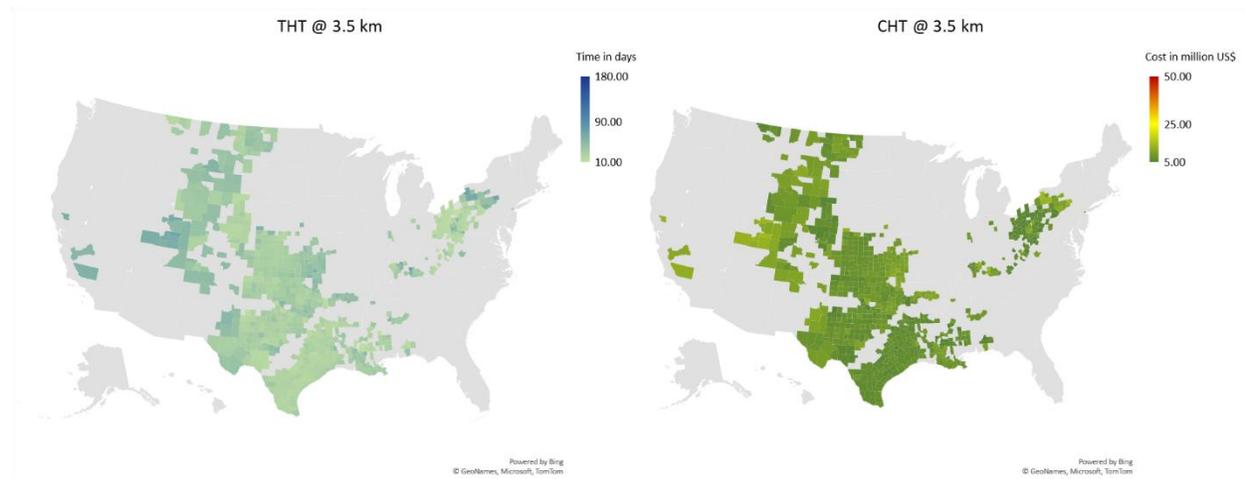

**Figure 8. Time-to-Hit-Temperature (THT) and Cost-to-Hit-Temperature (CHT) at 3.5 km (11483 ft)**

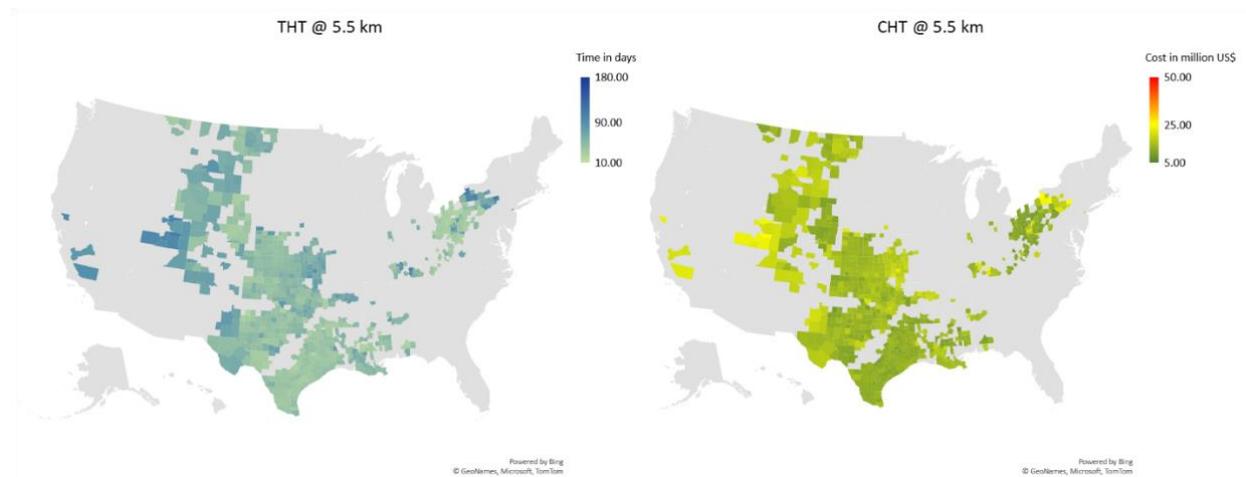

**Figure 9. Time-to-Hit-Temperature (THT) and Cost-to-Hit-Temperature (CHT) at 5.5 km (18045 ft)**

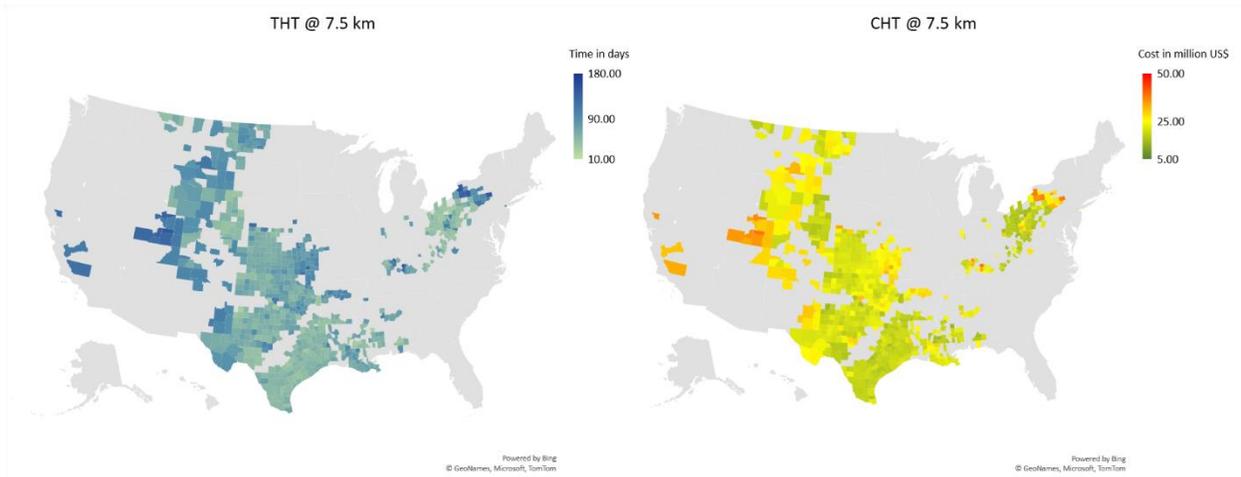

**Figure 10. Time-to-Hit-Temperature (THT) and Cost-to-Hit-Temperature, (CHT) at 7.5 km (24606 ft)**

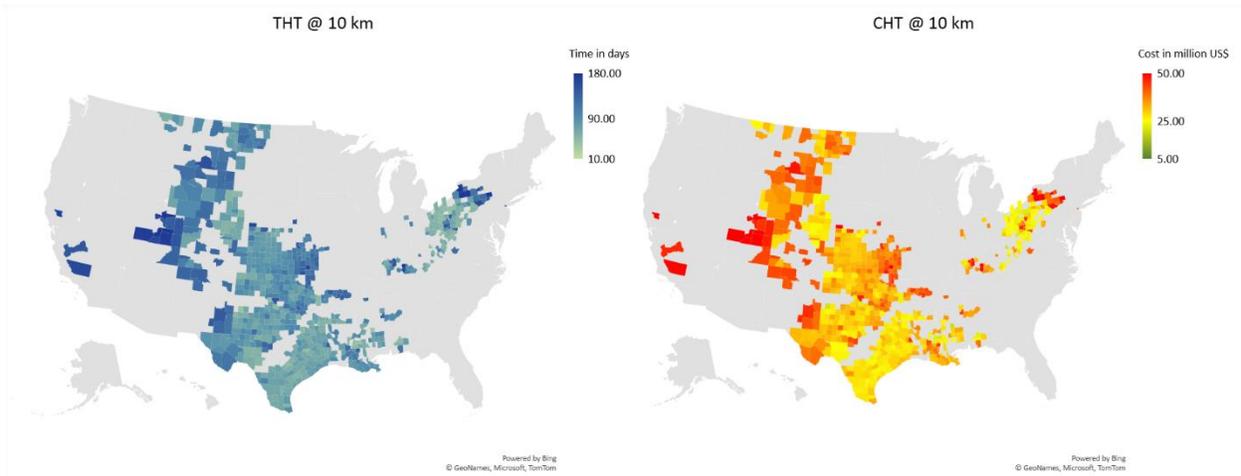

**Figure 11. Time-to-Hit-Temperature (THT) and Cost-to-Hit-Temperature, (CHT) at 10 km (32808 ft)**

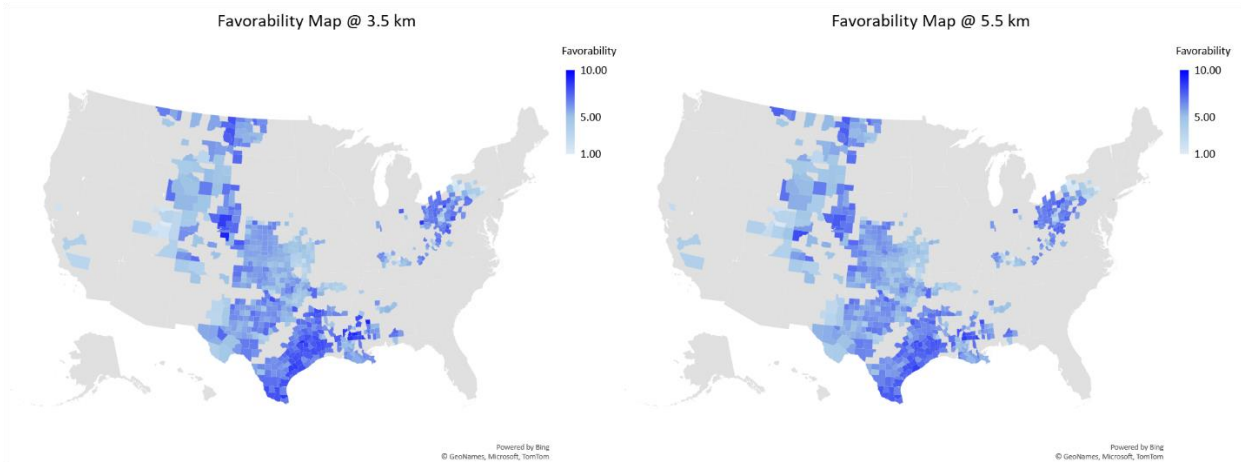

**Figure 12. Favorability Maps at 3.5 km (11483 ft) and 5.5 km (18045 ft)**

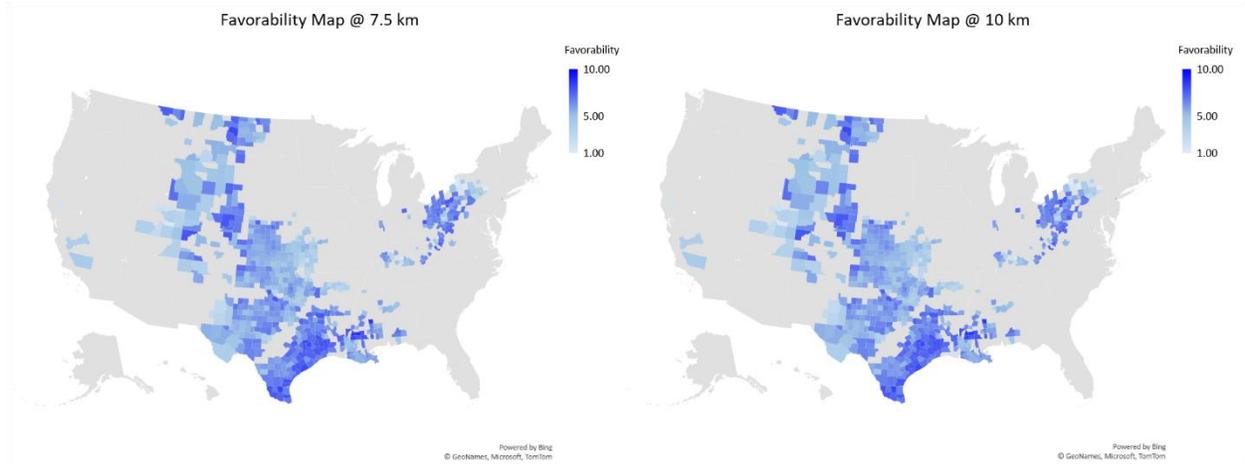

**Figure 13. Favorability Maps at 7.5 km (24606 ft) and 10 km (32808 ft)**

Further analysis and research are crucial to address the limitations and uncertainties associated with drilling and completion time data in geothermal projects. The reported data on drilling and completion time may be skewed towards the higher end due to various factors such as delays in completion operations, availability of stimulation services, or economic fluctuations affecting the timing of operations.

Integration of uncertainty in the data source is essential to provide a more comprehensive understanding of the drilling and completion process. Research efforts should focus on incorporating the inherent uncertainties and variability into the analysis, considering factors such as geological variations, rock properties, and unexpected challenges encountered during drilling operations. Accurately predicting well drilling and completion time and cost for deeper wells is indeed complex and challenging. The subsurface geology and rock properties become more uncertain and difficult to predict as drilling depth increases. This can lead to unexpected delays and increased costs associated with encountering challenging geological formations.

Moreover, the specialized equipment and techniques required for drilling deeper wells can be more expensive to operate and maintain. The unpredictable nature of the drilling process, including encountering unexpected obstacles or having to adapt drilling techniques to accommodate unexpected geologic features, further adds to the challenges of predicting drilling and completion time and cost accurately. Investigating the challenges related to geological variability, rock properties, and the potential risks and uncertainties associated with drilling deeper wells is crucial.

## 4. Conclusion

In conclusion, a comprehensive analysis was conducted to assess the feasibility of Enhanced Geothermal Systems (EGS) development in sedimentary basins across the United States. The study involved examining drilling and completion time, well cost, and subsurface temperature data to generate time to hit temperature (THT), cost to hit temperature (CHT), and favorability maps at different depths. These maps provide valuable insights into the geologic and economic factors contributing to favorable conditions for EGS development.

The analysis revealed potential opportunities for deep EGS applications in regions such as south to south-east Texas, south Louisiana, north-east Montana, west North Dakota, central to western

Colorado, and eastern Utah. These areas exhibit favorable temperatures and relatively faster and cost-effective drilling operations. These findings highlight the investment potential of drilling deeper wells, as they offer higher geothermal energy production in the long term. However, it is important to acknowledge the challenges associated with accurately predicting drilling and completion time and cost for deeper wells, considering uncertainties in subsurface geology and drilling conditions. Furthermore, this study emphasizes the need for further research to address limitations in data collection and integration, as well as to account for the uncertainty generated by various data sources.

Overall, the findings of this study contribute to the body of knowledge surrounding geothermal energy systems and provide valuable insights for policymakers, investors, and stakeholders in identifying the most favorable locations for EGS development. Improved understanding of geological variability and enhanced prediction accuracy will enhance decision-making processes for EGS development and facilitate more informed investment decisions, promoting sustainable and renewable energy solutions for the future.

## Acknowledgement

This material is based upon work supported by the U.S. Department of Energy's Office of Energy Efficiency and Renewable Energy (EERE) under the Geothermal Technologies Office Award Number DE-EE0009962. Additional financial support was provided by the University of Oklahoma Libraries' Open Access Fund to publish this work.

## Full Legal Disclaimer

This report was prepared as an account of work sponsored by an agency of the United States Government. Neither the United States Government nor any agency thereof, nor any of their employees, makes any warranty, express or implied, or assumes any legal liability or responsibility for the accuracy, completeness, or usefulness of any information, apparatus, product, or process disclosed, or represents that its use would not infringe privately owned rights. Reference herein to any specific commercial product, process, or service by trade name, trademark, manufacturer, or otherwise does not necessarily constitute or imply its endorsement, recommendation, or favoring by the United States Government or any agency thereof. The views and opinions of authors expressed herein do not necessarily state or reflect those of the United States Government or any agency thereof.